\documentclass[iop]{emulateapj}
 \usepackage{apjfonts}

\usepackage{amsmath}

\usepackage{floatflt, epsfig}

\def\sun{\hbox{$\odot$}}
\newcommand{\msun}{\!\,{\rm M}_\odot}

\def \spose#1{\hbox  to 0pt{#1\hss}}  
\def \lta{\mathrel{\spose{\lower 3pt\hbox{$\sim$}}\raise  2.0pt\hbox{$<$}}}
\def \gta{\mathrel{\spose{\lower  3pt\hbox{$\sim$}}\raise 2.0pt\hbox{$>$}}}

\shorttitle{Cosmic Variance Cookbook}
\shortauthors{Moster et al.}
\submitted{The Astrophysical Journal, submitted}

\begin{document}

\title{A Cosmic Variance Cookbook}

\author{Benjamin P. Moster\altaffilmark{1}, Rachel S. Somerville\altaffilmark{2,3}, Jeffrey A. Newman\altaffilmark{4} and Hans-Walter Rix\altaffilmark{1}}

\altaffiltext{1}{Max-Planck-Institut f\"ur Astronomie, K\"onigstuhl 17, 69117 Heidelberg, Germany;
   moster@mpia.de, rix@mpia.de.}

\altaffiltext{2}{Space Telescope Science Institute, 3700 San Martin Drive, Baltimore MD 21218; somerville@stsci.edu.}

\altaffiltext{3}{Department of Physics and Astronomy, The Johns Hopkins University, 3400 N. Charles St., Baltimore, MD 21218}

\altaffiltext{4}{Department of Physics and Astronomy, University of Pittsburgh,
   3941 O'Hara Street, Pittsburgh, PA 15260; janewman@pitt.edu.}

\begin{abstract}

Deep pencil beam surveys ($<1{\rm deg}^2$) are of fundamental
importance for studying the high-redshift universe. However,
inferences about galaxy population properties (e.g. the abundance of
objects) are in practice limited by \lq cosmic variance\rq. This is
the uncertainty in observational estimates of the number density of
galaxies arising from the underlying large-scale density
fluctuations. This source of uncertainty can be significant,
especially for surveys which cover only small areas and for massive
high-redshift galaxies. Cosmic variance for a given galaxy population
can be determined using predictions from cold dark matter theory and
the galaxy bias.
In this paper we provide tools for experiment design and
interpretation. For a given survey geometry we present the cosmic
variance of dark matter as a function of mean redshift $\bar{z}$ and
redshift bin size $\Delta z$. Using a halo occupation model to predict
galaxy clustering, we derive the galaxy bias as a function of mean
redshift for galaxy samples of a given stellar mass range. In the
linear regime, the cosmic variance of these galaxy samples is the
product of the galaxy bias and the dark matter cosmic variance. We
present a simple recipe using a fitting function to compute cosmic
variance as a function of the angular dimensions of the field,
$\bar{z}$, $\Delta z$ and stellar mass $m_*$. We also provide tabulated
values and a software tool.
We find that for GOODS at $\bar{z}=2$ and with $\Delta z=0.5$ the
relative cosmic variance of galaxies with $m_*>10^{11}\msun$ is $\sim
38\%$, while it is $\sim 27\%$ for GEMS and $\sim 12\%$ for
COSMOS. For galaxies of $m_*\sim10^{10}\msun$ the relative cosmic
variance is $\sim 19\%$ for GOODS, $\sim 13\%$ for GEMS and $\sim 6\%$
for COSMOS.  This implies that cosmic variance is a significant source
of uncertainty at $\bar{z}=2$ for small fields and massive galaxies,
while for larger fields and intermediate mass galaxies cosmic variance
is less serious.

\end{abstract}


\section{Introduction} \label{s:introduction}

Over the last decade, deep \lq pencil beam\rq~surveys of high-redshift
galaxies have been the observational basis for studying the processes
that drive galaxy formation and evolution. For a given amount of
observing time, such surveys have to trade off imaging area and
imaging depth: among HST imaging surveys, GEMS \citep{gems}, AEGIS
\citep{aegis} and COSMOS \citep{cosmos} are examples that aim at
comparatively wide fields to observe a large sample of galaxies.
Alternatively, the Hubble Deep Field \citep[HDF]{hdf} and the Ultra
Deep Field \citep[UDF]{udf} are examples of extremely deep surveys in
small areas, aimed at detecting faint galaxies (which can be either of
low mass or at high redshift).

One of the most fundamental properties of galaxy sub-populations at
any epoch is their number density.  However, observational estimates
of galaxy number densities in finite volumes are subject to
uncertainty due to \textit{cosmic variance}, arising from underlying
large-scale density fluctuations and leading to uncertainties in
excess of na\"ive Poisson errors. Note that this source of uncertainty
is referred to as \lq sample variance\rq~ in other branches of
cosmology. For sampling volumes much larger than the typical
clustering scale of the observed objects, cosmic variance is not
significant. However, many important existing surveys have a sampling
volume that is small enough that cosmic variance may dominate the
uncertainties. This may be particularly true at high redshift, where
galaxies are expected to be much more strongly clustered than dark
matter \citep{kauffmann1999,baugh1999,coil2004,moster2009}. Still,
many published quantities which are based on number density
(e.g. luminosity functions, stellar mass functions, etc.) are quoted
with error budgets that do not properly account for cosmic
variance. As shown by \citet{trenti2008}, the normalization and the
slope of high-redshift luminosity functions can be affected by cosmic
variance errors.

Sound estimates of cosmic variance are helpful both for analyzing
existing observational data and for designing future surveys.  For a
detailed and general treatment of the cosmic error, we refer to
\citet{szapudi1999}, \citet{colombi2000} and \citet{szapudi2000}. Our
goal here is to provide a simple recipe to derive cosmic variance for
galaxy samples selected by their stellar mass $m_*$, for a field of
specified angular dimensions $\alpha_1$ and $\alpha_2$, at a mean
redshift $\bar{z}$ and for a redshift bin size of $\Delta z$. Several
previous works have provided estimates of cosmic variance for specific
sets of assumptions. For example, \citet{newman2002} presented
estimates of the cosmic variance as a function of field area and axis
ratio for the redshift range $0.7<z<1.5$, relevant to the DEEP2
redshift survey, but did not account for galaxy bias (the clustering
amplitude of galaxies relative to dark matter; see
Eqn.\ref{eqn:bias}). \citet{somerville2004} provided predictions that
could be used to estimate cosmic variance as a function of mean
redshift and survey volume, using the number density of the population
to estimate the bias, assuming one galaxy per halo. However,
\citet{somerville2004} used the approximation of spherical (rather
than rectangular or pencil beam) volumes. \citet{trenti2008} presented
estimates of cosmic variance for pencil beam geometries, again using
the number density of the population to estimate the host halo mass
and hence the bias.

Many galaxy samples in observational studies, however, are selected by
their stellar mass, and stellar mass is one of the fundamental
properties of galaxies. Galaxy clustering and bias are strong
functions of stellar mass, making it extremely important to account
for the mass dependence in cosmic variance estimates. Moreover, one
does not necessarily know the number density of a population of
interest a priori. We therefore take a different approach here,
providing estimates of cosmic variance explicitly as a function of
stellar mass and redshift. To do this, we make use of the results
previously presented in \citet[][M09]{moster2009}. We used a Halo
Occupation Distribution (HOD) model to empirically establish the
relationship between stellar mass and dark matter halo (or sub-halo)
mass at different redshifts, and then used a dissipationless N-body
simulation to compute the galaxy bias as a function of stellar mass
and redshift. We presented fitting functions for this quantity in
M09. We combine these with estimates of the cosmic variance for the
dark matter for rectangular cells as described in \citet{newman2002}.

The paper is organized as follows: in Section \ref{sec:method} we
describe how to define and how to compute cosmic variance for a galaxy
population, based on the underlying cold dark matter (CDM) theory, a
model for galaxy bias and for a combination of separate survey fields.
In Section \ref{sec:recipe} we spell out a recipe to compute cosmic
variance for a pencil beam geometry in four steps: (1) select survey
geometry, (2) choose mean redshift and redshift bin size, (3)
determine stellar mass interval and (4) compute cosmic variance. We
also apply this recipe to four existing surveys at a redshift of $z=2$
and $z=3.5$. In Section \ref{sec:properties} we compare cosmic
variance for different galaxy samples at different redshifts and
different survey geometries. Finally, we summarize our methods and
conclusions in Section \ref{sec:conclusions}.

Throughout, we assume cosmological parameters consistent with results
from WMAP-3 \citep{spergel2007} for a flat $\Lambda{\rm CDM}$
cosmological model: matter density $\Omega_m=0.26$, cosmological
constant $\Omega_\Lambda=0.74$, Hubble parameter $H_0 = 72
{\rm~km~s^{-1}~Mpc^{-1}}$, fluctuation amplitude $\sigma_8=0.77$ and
primordial power-spectrum $n_s=0.95$. All stellar masses are
calibrated to a \citet{kroupa2001} initial mass function.

\section{Method}
\label{sec:method}

The mean $\langle N\rangle$ and the variance $\langle
N^2\rangle$-$\langle N\rangle^2$ are given by the first and second
moments of the probability distribution $P_N(V)$, which describes the
probability of counting $N$ objects within a volume $V$. The relative
cosmic variance is defined as
\begin{equation}
\label{eqncosvar}
\sigma_v^2=\frac{\langle N^2\rangle - \langle N\rangle^2-\langle
  N\rangle}{\langle N\rangle^2}.
\end{equation}
The last term in the numerator represents the correction for Poisson
shot noise.  The second moment of the object counts is
\begin{equation}
\langle N^2\rangle = \langle N\rangle^2+\langle N\rangle+
\frac{\langle N\rangle^2}{V^2} \int_V{\rm d}V_{\rm a}{\rm d}V_{\rm b}
\; \xi(|\textbf{r}_{\rm a}-\textbf{r}_{\rm b}|)
\end{equation}
where $\xi$ is the two-point correlation function of the sample and
$V$ is the sample volume \citep[see][pg. 234]{peebles1980}.  Combining
this with Equation \eqref{eqncosvar}, the cosmic variance can be
written as
\begin{equation}
\label{cosvarint}
\sigma_v^2=\frac{1}{V^2} \int_V{\rm d}V_{\rm a}{\rm d}V_{\rm b} \; \xi(|\textbf{r}_{\rm a}-\textbf{r}_{\rm b}|).
\end{equation}

There are two approaches to evaluating this equation. The first is
applicable to populations with a known correlation function, when the
integral can be solved either analytically or numerically as a
function of cell radius or cell volume.  However, in practice, there
are a number of difficulties with this approach: while on small scales
in the nonlinear regime ($r<10-15~{\rm Mpc}$) the correlation function
can typically be approximated by a power law, it deviates on larger
scales (in the linear regime), from this form. Additionally, in many
cases of interest the correlation function is not known a priori.

The second approach to calculating the cosmic variance can be used if
the correlation function cannot be approximated by a power law or if
it is unknown. It makes use of the galaxy bias, which can be predicted
for a given galaxy population using halo occupation models.

We can substitute the galaxy correlation function $\xi_{gg}$ in
Equation \eqref{cosvarint} for the correlation function for dark
matter $\xi_{dm}$:
\begin{equation}
\xi_{gg}(r,m,z)=b^2(m_*,z)\; \xi_{dm}(r,z)\;,
\label{eqn:bias}
\end{equation}
where $b(m_*,z)$ is the galaxy bias, which depends both on redshift
$z$, and the stellar mass of the galaxies $m_*$. In general, the bias
is also a function of $r$, however, in the linear regime, we assume
that it is independent of scale. This assumption will be made
throughout the paper. Employing the bias in this manner yields
 \begin{eqnarray}
 \label{eqn:cvdm}
\sigma_v^2(m_*,z)&=&\frac{1}{V^2} \int_V{\rm d}V_{\rm a}{\rm d}V_{\rm b}\; b^2(m_*,z)\; \xi_{dm}(r_{\rm ab},z)\notag\\
&=&b^2(m_*,z)\; \frac{1}{V^2} \int_V{\rm d}V_{\rm a}{\rm d}V_{\rm b}\; \xi_{dm}(r_{\rm ab},z)\notag\\
&=&b^2(m_*,z)\; \sigma_{dm}^2(z)
\end{eqnarray}
The integral leading to $\sigma_{dm}$ can then be solved using linear
theory in CDM for any volume $V$ with a given geometry. The cosmic
variance for galaxies can thus be determined by multiplying the dark
matter cosmic variance at a given redshift with the galaxy bias for a
given stellar mass at that redshift.  For a pencil beam geometry the
volume element $V$ is set by the angular size of the survey and the
redshift bin size.

We now pursue this second approach, using the galaxy bias as a
function of redshift and stellar mass as computed by
\citet{moster2009}. In this paper the authors used a halo occupation
model to obtain a parameterized stellar-to-halo mass relation by
populating halos and subhalos in an $N$-body simulation with galaxies
and requiring that the observed stellar mass function be
reproduced. Using the halo positions obtained from the simulation,
they find that predictions for the galaxy correlation function are in
excellent agreement with observed clustering properties at low
redshift. The derived stellar-to-halo mass relation is finally used to
predict the stellar-mass-dependent galaxy correlation function and
bias at high redshift.

We calculate the cosmic variance for dark matter in each redshift bin
by solving the integral in Eqn~\eqref{cosvarint} for dark matter using
the code {\sc QUICKCV}\footnote[1]{{\sc QUICKCV} is available at
  http://astro.berkeley.edu/~jnewman/research} which is described in
\citet{newman2002}. The code computes the dark matter correlation
function using the transfer function given in \citet{bardeen1986} and
a window function which is 1 inside the volume $V$ and 0
elsewhere. The volume elements are determined by the angular
dimensions of the survey field and the redshift bin. Finally, the
cosmic variance for galaxies is computed using Equation
\eqref{eqn:cvdm}.

Once cosmic variance has been determined for a number of individual
survey fields that are widely separated on the sky, compared to the
correlation length, as is often the case (e.g. for the various HST
survey fields), one can easily derive the cosmic variance for the
combined sample. For the combination of fields $i$ the total volume is
\begin{equation}
V_{\rm tot}=\Sigma_i\;V_i\;,
\end{equation}
implying a cosmic variance for the total sample of
\begin{eqnarray}
\sigma_{\rm tot}^2 &=& \frac{1}{V_{\rm tot}^2} \int_V{\rm d}V_{\rm a}{\rm d}V_{\rm b} \; \xi(|\textbf{r}_{\rm a}-\textbf{r}_{\rm b}|)\notag\\
&=& \frac{1}{\left(\Sigma_i\;V_i\right)^2} \int_V{\rm d}(\Sigma_i\;V_i)_{\rm a}{\rm d}(\Sigma_j\;V_j)_{\rm b} \; \xi(|\textbf{r}_{i\rm a}-\textbf{r}_{j\rm b}|)\notag\\
&=& \frac{1}{\left(\Sigma_i\;V_i\right)^2}\;\Sigma_{ij} \int_V {\rm d}V_{i\rm a}{\rm d}V_{j\rm b} \; \xi(|\textbf{r}_{i\rm a}-\textbf{r}_{j\rm b}|)\notag\\
&=& \frac{1}{\left(\Sigma_i\;V_i\right)^2}\; \Sigma_i \int_V {\rm d}V_{i\rm a}{\rm d}V_{i\rm b} \; \xi(|\textbf{r}_{i\rm a}-\textbf{r}_{i\rm b}|)\notag\\
&=& \frac{\Sigma_i V_i^2 \sigma_i^2}{\left(\Sigma_i\;V_i\right)^2}\;.
\end{eqnarray}
where we have used
\begin{equation}
\xi(|\textbf{r}_{i\rm a}-\textbf{r}_{j\rm b}|) = 0 \; \text{for } i\neq j
\end{equation}
since the galaxies in separated fields are uncorrelated. For a pencil beam geometry (angular extend $\ll 1$ radian) this can be written as
\begin{equation}
\label{eqn:addfields}
\sigma_{\rm tot}^2 = \frac{\Sigma_i\;(\alpha_1\alpha_2)_i^2 \sigma_i^2}{\left[\Sigma_i\;(\alpha_1\alpha_2)_i\right]^2}\;
\end{equation}
where $\alpha_1$ and $\alpha_2$ are the angular dimensions of each
field. Note that this volume-weighted average is identical to an
inverse-variance-weighted average only in the range where the cosmic
variance scales with volume.

\section{Recipe for cosmic variance}
\label{sec:recipe}

In order to compute the cosmic variance for galaxies in a given volume
in the universe one has to fix five parameters. The size of the volume
is determined by two dimensions tangential to the line of sight and
one dimension parallel to the line of sight. The tangential dimensions
we use are the angular size of the field $\alpha_1$ and $\alpha_2$,
while the vertical dimension is set by the size of the redshift bin
under consideration $\Delta z$. The fourth parameter is the mean
redshift of the volume $\bar{z}$, since the cosmic variance depends on
the clustering strength which is a function of redshift. Moreover
$\bar{z}$ determines the length scale of the volume bins by setting
the conversion factor from angular dimensions and $\Delta z$ into
comoving Mpc.  In order to determine the cosmic variance for galaxies
one finally has to fix the stellar mass $m_*$ of the sample under
consideration, as the clustering strength and the bias depend strongly
on stellar mass.  Thus the five parameters which have to be set are
$\alpha_1$, $\alpha_2$, $\bar{z}$, $\Delta z$ and $m_*$.

\subsection{Survey Geometry}
\label{sec:survey}

The first step to calculate the cosmic variance is to fix the size of
the volume element under consideration. For now we set the size of the
redshift bin to a reference value of $\Delta z = 0.2$.  This is a
convenient bin size for low redshift while for higher redshift it is
rather small. However, this choice can be changed at any redshift (for
values $\Delta z \geq 0.05$) using the recipe presented in section
\ref{sec:redshiftbins}. For the tangential dimensions, we choose five
popular fields: the Hubble Ultra Deep Field \citep[UDF]{udf}, one
field of the Great Observatories Origins Deep Survey
\citep[GOODS]{goods}, the Extended Chandra Deep Field South (E-CDFS)
used in the Galaxy Evolution From Morphology And SEDs survey
\citep[GEMS]{gems}, the Extended Groth Strip (EGS) used in the
All-wavelength Extended Groth strip International Survey
\citep[AEGIS]{aegis} and the field used in the Cosmic Evolution Survey
\citep[COSMOS]{cosmos}.  The angular dimensions of the surveys and the
resulting observation areas are summarized in Table
\ref{t:surveyarea}. 
We note that due to the small scales for the UDF numerical
uncertainties in the integration of the correlation function are
likely to be high. As a result the values for cosmic variance in the
UDF computed here, especially for $z<0.5$ (where size of the volume
elements is very small) have considerable uncertainties. At these
scales, the assumption of linear biasing is also not accurate.  As a
consequence the cosmic variance for the UDF presented here is to be
treated with caution and is mainly included to demonstrate the large
cosmic variance in small fields.

\begin{deluxetable}{l r r r r r}
\tablecaption{Summary of the survey geometries}
\tablehead{
  \colhead{Survey} &
  \colhead{$N_{\rm field}$} &
  \colhead{$\alpha_1$\tablenotemark{a}} &
  \colhead{$\alpha_2$\tablenotemark{a}} &
  \colhead{${\rm area}_{\rm field}$\tablenotemark{b}} &
  \colhead{total area\tablenotemark{b}}
}
\startdata
UDF & 1 & 3.3 & 3.3 & 11 & 11\\
GOODS  & 2 & 10 & 16 & 160 & 320\\
GEMS  & 1 & 28 & 28 & 784 & 784\\
EGS  & 1 & 10 & 70 & 700 & 700\\
COSMOS  & 1 & 84 & 84 & 7056 & 7056\\
\enddata
\tablenotetext{a}{Angular dimension in arcmin}
\tablenotetext{b}{Areas in ${\rm arcmin}^2$}
\label{t:surveyarea}
\end{deluxetable}

\begin{figure}
\psfig{figure=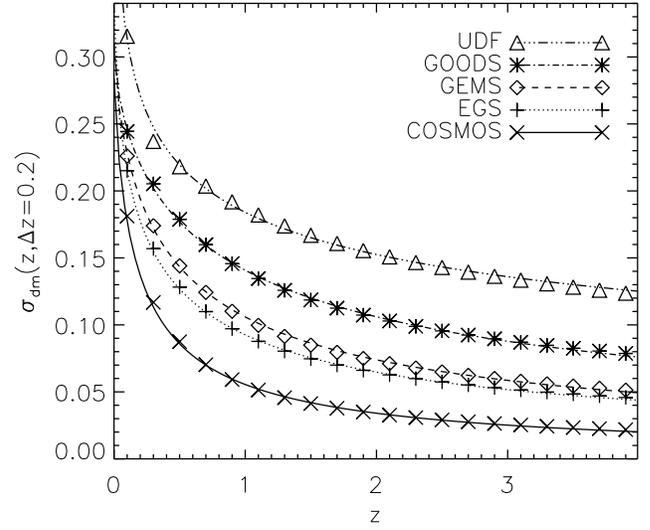,width=0.47\textwidth}
\caption{\scriptsize Cosmic variance as a function of mean redshift
  with a reference bin size of $\Delta z_{\rm ref}=0.2$. The different
  symbols correspond to different fields while the lines are fits to
  the symbols using Equation \eqref{eqn:meanredshift}.}
\label{fig:cv}
\end{figure}

\begin{deluxetable}{c c c c c c c c c}
\tablecaption{Cosmic variance for different surveys with common
  redshift bins. The first and second columns show the mean redshift
  and the redshift bin size, the third column gives the root comic
  variance for dark matter while the last six columns give the root
  cosmic variance for galaxies in the indicated stellar mass
  intervals.}

\startdata
\hline
\hline\\
~ & ~ & ~ & \multicolumn{6}{c}{$\sigma_{gg}$ for $\log(m_*/\msun)\pm0.25$}\\
$\bar{z}$ & $\Delta z$ & $\sigma_{dm}$ & ~ & ~ & ~ & ~ & ~\\
~&~&~&8.75&9.25&9.75&10.25&10.75&11.25\\

\cutinhead{UDF $3.3'\times3.3'$}\\
0.1 & 0.2 & 0.316 & 0.349 & 0.358 & 0.380 & 0.409 & 0.430 & 0.526\\
0.3 & 0.2 & 0.237 & 0.272 & 0.288 & 0.295 & 0.318 & 0.339 & 0.428\\
0.5 & 0.2 & 0.218 & 0.262 & 0.273 & 0.283 & 0.307 & 0.332 & 0.435\\
0.7 & 0.2 & 0.204 & 0.259 & 0.271 & 0.280 & 0.305 & 0.335 & 0.456\\
0.9 & 0.2 & 0.192 & 0.259 & 0.274 & 0.282 & 0.308 & 0.346 & 0.488\\
1.1 & 0.2 & 0.182 & 0.264 & 0.281 & 0.289 & 0.317 & 0.365 & 0.531\\
1.5 & 0.2 & 0.167 & 0.282 & 0.305 & 0.319 & 0.352 & 0.426 & 0.648\\
1.9 & 0.2 & 0.156 & 0.311 & 0.341 & 0.369 & 0.407 & 0.518 & 0.807\\
2.5 & 0.2 & 0.143 & 0.374 & 0.416 & 0.482 & 0.529 & 0.718 & 1.129\\
3.5 & 0.2 & 0.128 & 0.523 & 0.593 & 0.781 & 0.845 & 1.236 & 1.898\\

\cutinhead{GOODS $10'\times16'$}\\
0.1 & 0.2 & 0.244 & 0.270 & 0.277 & 0.294 & 0.317 & 0.333 & 0.407\\
0.3 & 0.2 & 0.205 & 0.236 & 0.243 & 0.255 & 0.276 & 0.293 & 0.371\\
0.5 & 0.2 & 0.179 & 0.215 & 0.224 & 0.232 & 0.252 & 0.272 & 0.357\\
0.7 & 0.2 & 0.160 & 0.203 & 0.213 & 0.220 & 0.239 & 0.263 & 0.358\\
0.9 & 0.2 & 0.146 & 0.197 & 0.208 & 0.214 & 0.234 & 0.263 & 0.371\\
1.1 & 0.2 & 0.135 & 0.195 & 0.207 & 0.214 & 0.235 & 0.270 & 0.392\\
1.5 & 0.2 & 0.119 & 0.200 & 0.217 & 0.227 & 0.250 & 0.303 & 0.460\\
1.9 & 0.2 & 0.107 & 0.215 & 0.235 & 0.255 & 0.281 & 0.357 & 0.557\\
2.5 & 0.2 & 0.096 & 0.250 & 0.278 & 0.322 & 0.354 & 0.480 & 0.754\\
3.5 & 0.2 & 0.083 & 0.336 & 0.382 & 0.503 & 0.544 & 0.795 & 1.221\\

\cutinhead{GEMS $28'\times28'$}\\
0.1 & 0.2 & 0.226 & 0.250 & 0.256 & 0.272 & 0.293 & 0.308 & 0.377\\
0.3 & 0.2 & 0.174 & 0.200 & 0.206 & 0.216 & 0.234 & 0.249 & 0.315\\
0.5 & 0.2 & 0.144 & 0.173 & 0.180 & 0.187 & 0.203 & 0.219 & 0.288\\
0.7 & 0.2 & 0.124 & 0.158 & 0.165 & 0.171 & 0.186 & 0.204 & 0.279\\
0.9 & 0.2 & 0.110 & 0.149 & 0.157 & 0.162 & 0.177 & 0.199 & 0.280\\
1.1 & 0.2 & 0.100 & 0.144 & 0.154 & 0.158 & 0.174 & 0.200 & 0.291\\
1.5 & 0.2 & 0.085 & 0.143 & 0.155 & 0.162 & 0.179 & 0.217 & 0.329\\
1.9 & 0.2 & 0.075 & 0.150 & 0.164 & 0.178 & 0.196 & 0.250 & 0.389\\
2.5 & 0.2 & 0.065 & 0.170 & 0.189 & 0.219 & 0.240 & 0.326 & 0.512\\
3.5 & 0.2 & 0.054 & 0.221 & 0.251 & 0.331 & 0.358 & 0.523 & 0.804\\

\cutinhead{EGS $10'\times70'$}\\
0.1 & 0.2 & 0.215 & 0.238 & 0.244 & 0.259 & 0.279 & 0.293 & 0.358\\
0.3 & 0.2 & 0.157 & 0.180 & 0.186 & 0.195 & 0.211 & 0.224 & 0.284\\
0.5 & 0.2 & 0.128 & 0.154 & 0.160 & 0.167 & 0.181 & 0.195 & 0.256\\
0.7 & 0.2 & 0.120 & 0.140 & 0.146 & 0.151 & 0.164 & 0.181 & 0.246\\
0.9 & 0.2 & 0.097 & 0.131 & 0.139 & 0.143 & 0.156 & 0.175 & 0.247\\
1.1 & 0.2 & 0.088 & 0.127 & 0.135 & 0.139 & 0.153 & 0.176 & 0.256\\
1.5 & 0.2 & 0.075 & 0.126 & 0.136 & 0.143 & 0.157 & 0.191 & 0.290\\
1.9 & 0.2 & 0.066 & 0.132 & 0.145 & 0.157 & 0.173 & 0.220 & 0.343\\
2.5 & 0.2 & 0.057 & 0.150 & 0.167 & 0.194 & 0.213 & 0.288 & 0.453\\
3.5 & 0.2 & 0.048 & 0.197 & 0.224 & 0.295 & 0.319 & 0.466 & 0.716\\

\cutinhead{COSMOS $84'\times84'$}\\
0.1 & 0.2 & 0.181 & 0.200 & 0.205 & 0.218 & 0.235 & 0.247 & 0.302\\
0.3 & 0.2 & 0.117 & 0.134 & 0.138 & 0.145 & 0.157 & 0.167 & 0.211\\
0.5 & 0.2 & 0.088 & 0.105 & 0.109 & 0.114 & 0.123 & 0.133 & 0.175\\
0.7 & 0.2 & 0.070 & 0.089 & 0.094 & 0.097 & 0.105 & 0.116 & 0.158\\
0.9 & 0.2 & 0.059 & 0.080 & 0.084 & 0.087 & 0.095 & 0.107 & 0.151\\
1.1 & 0.2 & 0.051 & 0.074 & 0.079 & 0.083 & 0.090 & 0.103 & 0.150\\
1.5 & 0.2 & 0.041 & 0.070 & 0.075 & 0.079 & 0.087 & 0.105 & 0.160\\
1.9 & 0.2 & 0.035 & 0.070 & 0.077 & 0.083 & 0.092 & 0.117 & 0.182\\
2.5 & 0.2 & 0.029 & 0.076 & 0.085 & 0.098 & 0.108 & 0.146 & 0.230\\
3.5 & 0.2 & 0.023 & 0.095 & 0.108 & 0.142 & 0.154 & 0.225 & 0.346\\

\enddata
\label{t:cvsurveylin}
\end{deluxetable} 

\newpage

\subsection{Redshift bins}
\label{sec:redshiftbins}

The next step is to calculate the cosmic variance for dark matter for
a set of mean redshifts $\bar{z}$. We do this using the angular
dimensions of the five surveys presented in section \ref{sec:survey}
and the reference redshift bin size of $\Delta z = 0.2$. The cosmic
variance is then computed for a set of mean redshifts. The resulting
values of the root cosmic variance for dark matter are given in the
third column of Table \ref{t:cvsurveylin} and are plotted in Figure
\ref{fig:cv} (symbols). To make it possible to obtain the cosmic
variance for mean redshifts other than these tabulated values, we
introduce a fitting function:

\begin{eqnarray}
\label{eqn:meanredshift}
\sigma_{dm}(\bar{z},\Delta z=0.2) = \frac{\sigma_a}{\bar{z}^\beta+\sigma_b}
\end{eqnarray}

\begin{deluxetable}{l c c c c c c}
\tablecaption{Fitting parameters for different surveys}
\tablehead{
  \colhead{Survey} &
  \colhead{~} &
  \colhead{$\sigma_a$} &
  \colhead{~} &
  \colhead{$\sigma_b$} &
  \colhead{~} &
  \colhead{$\beta$}
}
\startdata
UDF & & 0.251 & & 0.364 & & 0.358\\
GOODS  & & 0.261 & & 0.854 & & 0.684\\
GEMS  & & 0.161 & & 0.520 & & 0.729\\
EGS  & & 0.128 & & 0.383 & & 0.673\\
COSMOS  & & 0.069 & & 0.234 & & 0.834\\
\enddata
\label{tab:meanredshift}
\end{deluxetable} 

The parameters $\sigma_a$, $\sigma_b$ and $\beta$ depend on the
angular dimensions of the field and are given in Table
\ref{tab:meanredshift}. The lines in Figure \ref{fig:cv} show the
fitting function for the five surveys and agree very well with the
computed values for the redshift bins. Using Equation
\eqref{eqn:meanredshift} for any field we are thus able to calculate
the dark matter cosmic variance as a function of mean redshift.

\begin{figure}
\psfig{figure=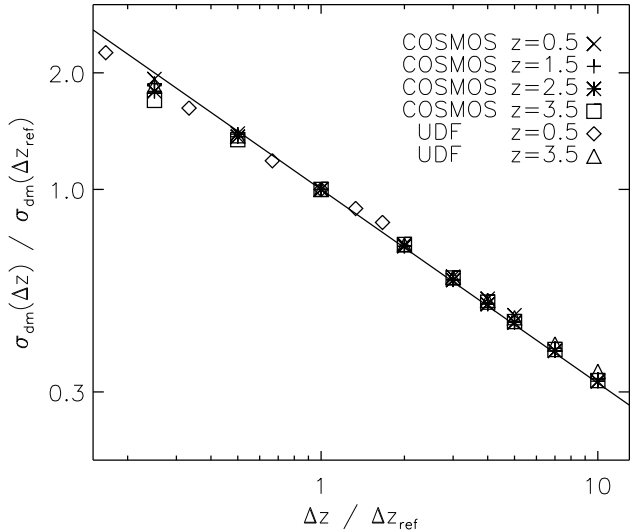,width=0.47\textwidth}
\caption{\scriptsize Cosmic variance for different redshift bin sizes
  normalized to the reference size $\Delta z_{\rm ref}=0.2$. The
  different symbols correspond to different fields and mean redshifts
  and the solid line to $\sqrt{\Delta z_{\rm ref} / \Delta z}$. The
  dependence on $\Delta z$ is similar for all geometries and
  redshifts.}
\label{fig:deltaz}
\end{figure}

Up until now we have assumed a reference redshift bin size of $\Delta
z_{\rm ref}=0.2$ in order to determine the size of the volume element
under consideration. However, especially at higher redshift, larger
bin sizes are desirable. We thus investigate how the cosmic variance
depends on $\Delta z$. For this we plot $\sigma_{\rm dm}(\Delta z) /
\sigma_{\rm dm}(\Delta z_{\rm ref})$ as a function of $\Delta z/\Delta
z_{ref}$ for different fields and mean redshifts in Figure
\ref{fig:deltaz}.  We find that independent of angular dimensions and
mean redshift the cosmic variance has the same dependance on the
redshift bin size:
\begin{eqnarray}
\frac{\sigma_{\rm dm}(\Delta z)}{\sigma_{\rm dm}(\Delta z_{\rm ref})} = \left(\frac{\Delta z}{\Delta z_{ref}}\right)^{-0.5}
\end{eqnarray}
This means that for a reference redshift bin size of $\Delta z_{\rm
  ref}=0.2$ as assumed in Equation \eqref{eqn:meanredshift} the root
cosmic variance can be calculated for a different $\Delta z$ as:
\begin{eqnarray}
\label{eqn:deltaz}
\sigma_{\rm dm}(\Delta z,\bar{z})=\sigma_{\rm dm}(\Delta z=0.2,\bar{z}) \; \sqrt{\frac{0.2}{\Delta z}}
\end{eqnarray}
There is a simple picture that explains the dependence on $\Delta
z$. If we divide a redshift bin of size $\Delta z$ into $N$ redshift
bins with size $\Delta z_i$ then $\Delta z = N \Delta z_i$. Assuming
that all these redshift bins are uncorrelated, we can invoke Equation
\eqref{eqn:addfields} which yields $\sigma(\Delta z)=\sigma(\Delta
z_i)/\sqrt{N}$. Using $N=\Delta z/\Delta z_i$ we finally get
$\sigma(\Delta z)=\sigma(\Delta z_i) \times \sqrt{\Delta z_i/\Delta
  z}$. It is interesting that the simplified assumption of
uncorrelated volume elements yields a relation that fits so well to
the results calculated for the total redshift bin, as seen in Figure
\ref{fig:deltaz}. However, we note that this approximation is only
valid for redshift bin sizes $\Delta z>0.05$. For smaller bin sizes we
refer to the more accurate model presented in section 3.5 of
\citet{newman2008}.

\subsection{Stellar mass dependence}
\label{sec:stellarmass}

The last step to derive the cosmic variance for a given sample of
galaxies is to apply the galaxy bias.  As we have shown in Equation
\eqref{eqn:cvdm}, in the linear-biasing limit, the cosmic variance for
a sample of galaxies with a given stellar mass is the product of the
squared bias and the dark matter cosmic variance. The stellar mass
dependent galaxy bias has been derived in \citet{moster2009}.  The
authors present the redshift dependence of the bias using
parameterized functions of the form:
\begin{equation}
\label{eqn:bias}
b(m_*,\bar z)=b_0 (\bar z+1)^{b_1}+b_2 \quad
\end{equation}
with the parameters $b_0$, $b_1$, and $b_2$ which are given in Table
\ref{tab:biasfit} for six stellar mass bins and six stellar mass
thresholds. In order to compute the root cosmic variance for galaxies
of that mass, we have to multiply the bias and the root cosmic
variance of dark matter for the same redshift. This has been done for
the five surveys presented in Table \ref{t:cvsurveylin}: the cosmic
variance for galaxies of mass $m_*$ is given in columns 4-9 for the
reference redshift bin size of $\Delta z=0.2$. Cosmic variance for
massive galaxies is always larger than that of low mass galaxies,
since the galaxy bias increases with increasing stellar mass.

\begin{deluxetable}{lccc}
\tablecaption{Galaxy bias fit parameters}
\tablehead{
  \colhead{$\log m_g$} &
  \colhead{$b_0$} &
  \colhead{$b_1$} &
  \colhead{$b_2$}
}
\startdata
~~8.5~-~~~9.0 & 0.062 $\pm$ 0.017 & 2.59 $\pm$ 0.18 & 1.025 $\pm$ 0.062\\
~~9.0~-~~~9.5 & 0.074 $\pm$ 0.008 & 2.58 $\pm$ 0.26 & 1.039 $\pm$ 0.028\\
~~9.5~-~10.0 & 0.042 $\pm$ 0.003 & 3.17 $\pm$ 0.05 & 1.147 $\pm$ 0.021\\
10.0~-~10.5 & 0.053 $\pm$ 0.014 & 3.07 $\pm$ 0.17 & 1.225 $\pm$ 0.077\\
10.5~-~11.0 & 0.069 $\pm$ 0.014 & 3.19 $\pm$ 0.13 & 1.269 $\pm$ 0.087\\
11.0~-~11.5 & 0.173 $\pm$ 0.035 & 2.89 $\pm$ 0.20 & 1.438 $\pm$ 0.061\\
$~~>~~8.5$ & 0.063 $\pm$ 0.008 & 2.62 $\pm$ 0.08 & 1.104 $\pm$ 0.028\\
$~~>~~9.0$ & 0.085 $\pm$ 0.008 & 2.50 $\pm$ 0.06 & 1.098 $\pm$ 0.064\\
$~~>~~9.5$ & 0.058 $\pm$ 0.005 & 2.96 $\pm$ 0.06 & 1.192 $\pm$ 0.021\\
$~~>~10.0$ & 0.072 $\pm$ 0.013 & 2.90 $\pm$ 0.13 & 1.257 $\pm$ 0.051\\
$~~>~10.5$ & 0.093 $\pm$ 0.015 & 3.02 $\pm$ 0.11 & 1.332 $\pm$ 0.054\\
$~~>~11.0$ & 0.185 $\pm$ 0.032 & 2.86 $\pm$ 0.25 & 1.448 $\pm$ 0.098\\
\enddata
\tablecomments{All quoted masses are in units of $\msun$}
\label{tab:biasfit}
\end{deluxetable}

\subsection{Cookbook for Cosmic Variance}
\label{sec:computecv}
Finally we can summarize our recipe to derive the cosmic variance for
a particular survey:

\begin{description}

\item[Step 1)] Choose the survey (field) in Table \ref{t:surveyarea}
  that is closest to the field you are using.

\item[Step 2)] Choose a mean redshift $\bar z$ and a redshift bin size
  $\Delta z$.

\item[Step 3)] Choose the stellar mass range of your galaxy sample from
  Table \ref{tab:biasfit}.

\item[Step 4a)] Calculate the dark matter root cosmic variance
  $\sigma_{dm}(\bar{z},\Delta z=0.2)$ using Equation
  \eqref{eqn:meanredshift} and the parameters for your survey as given
  in Table \ref{tab:meanredshift}.

\item[Step 4b)] Calculate the galaxy bias $b(m_*,\bar z)$ using
  Equation \eqref{eqn:bias} and the parameters for your stellar mass
  bin or threshold as given in Table \ref{tab:biasfit}.

\item[Step 4c)] Compute the root cosmic variance for galaxies:
$$
\sigma_{\rm gg}(m_*,\bar z,\Delta z)=b(m_*,\bar z) \; \sigma_{dm}(\bar{z},\Delta z=0.2) \; \sqrt{\frac{0.2}{\Delta z}}
$$

\item[Step 4d)] Combine widely separated fields using Equation
  \eqref{eqn:addfields}.

\end{description}

This recipe allows for easy computation of cosmic variance for the
representative surveys we have presented.  To enable readers to
compute cosmic variance for different survey areas or geometries, we
provide a software tool as an online supplement
\footnote[2]{IDL code available at http://www.mpia.de/homes/moster/research}.
For specified angular dimensions and redshift bins the code computes
the cosmic variance for dark matter and galaxies in six stellar mass bins.

\subsection{Examples}

Having derived the recipe to determine the cosmic variance, we can
apply it to the five example surveys. We first give an overview of the
surveys.  The UDF is an exposure of a square field in the southern sky
covering a total area of 11${~\rm arcmin}^2$ with the Advanced Camera
for Surveys (ACS) on the Hubble Space Telescope (HST). To date it is
the deepest image of the universe with uniform limiting magnitudes $m
\sim 29$ for point sources and contains at least 10,000
objects. However, of the five surveys studied here, the UDF is the
smallest.  The GOODS project covers two fields, the Hubble Deep Field
North (HDFN) and the Chandra Deep Field South (CDFS) which both have
similar dimensions of $\approx 10\times16 {~\rm arcmin}^2$ and are so
widely separated as to be uncorrelated. Since for widely separated
fields the cosmic variance goes as $1/N_{\rm field}$, the total
variance for both fields decreases by a factor of 2.  The GEMS survey
has imaged the E-CDFS, a square area of $784{~\rm arcmin}^2$, with the
ACS on the HST. The E-CDFS is centered on the CDFS and contains
roughly 10,000 galaxies down to a depth of 24th magnitude in the
R-band. AEGIS is a multi-wavelength, deep, wide-field, photometric and
spectroscopic survey in the Extended Groth Strip (EGS) area. While the
total area is similar to that of GEMS, the field is a long strip with
dimensions of $\approx 10\times70 {~\rm arcmin}^2$. The widest survey
for which we compute cosmic variance is COSMOS which covers an
equatorial square field of $2{~\rm deg}^2$ with imaging by most of the
major space-based telescopes and several large ground-based
telescopes. The magnitude limit in the I-band is $I \sim 27$.

If we apply this recipe to the UDF, GOODS, GEMS, AEGIS and COSMOS at
$\bar{z}=2$ and $\Delta z=0.5$, the relative cosmic variance for
massive galaxies with $m_*>10^{11}\msun$ is $\sim 55\%$ for the UDF,
while it is $\sim 38\%$ for GOODS, $\sim 27\%$ for GEMS, $\sim 23\%$
for AEGIS and $\sim 12\%$ for COSMOS.  For galaxies of intermediate
mass ($m_*\sim10^{10}\msun$) the relative cosmic variance is $\sim
27\%$ for the UDF, $\sim 19\%$ for GOODS, $\sim 13\%$ for GEMS, $\sim
11\%$ for AEGIS and $\sim 6\%$ for COSMOS.  This implies that at
$\bar{z}=2$ cosmic variance is a significant source of uncertainty for
small fields and massive galaxies. However, for larger fields and
intermediate mass galaxies cosmic variance is less serious.

At a higher redshift of $\bar{z}=3.5$ and $\Delta z=0.5$ the relative
cosmic variance is higher, since the galaxy bias is much larger. For
massive galaxies with $m_*>10^{11}\msun$, we find a relative cosmic
variance of $\sim 124\%$ for the UDF, $\sim 78\%$ for GOODS, $\sim
51\%$ for GEMS, $\sim 45\%$ for AEGIS and $\sim 21\%$ for COSMOS.  For
intermediate mass galaxies with $m_*\sim10^{10}\msun$ the relative
cosmic variance is $\sim 54\%$ for the UDF, $\sim 34\%$ for GOODS,
$\sim 22\%$ for GEMS, $\sim 20\%$ for AEGIS and $\sim 9\%$ for COSMOS.
This shows that at $\bar{z}=3.5$ cosmic variance can be serious in
small fields even for intermediate mass galaxies. For massive galaxies
even the widest survey (COSMOS) has considerable uncertainties due to
cosmic variance.

\section{Scaling of Cosmic Variance}
\label{sec:properties}

It is not very useful to compare the cosmic variance at different
redshifts for a fixed bin size, because the volumes are not the
same. Since cosmic variance strongly depends on the volume of the
sample under consideration, we choose redshift bins such that all
redshift bins have the same volume. This allows us to directly compare
the cosmic variance at different redshifts. The results of this
analysis are given in Table \ref{t:cvsurveyvol} for the GOODS, GEMS
and COSMOS surveys.

\begin{figure}
\psfig{figure=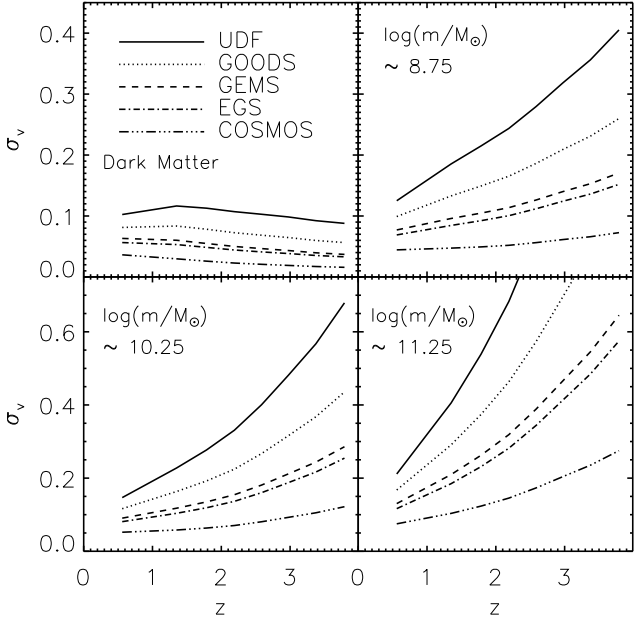,width=0.47\textwidth}
\caption{\scriptsize The root cosmic variance as a function of
  redshift with redshift bins of constant volume. The four panels are
  for dark matter and galaxies in three stellar mass bins.  The lines
  are for different surveys.}
\label{fig:cvsurvey}
\end{figure}

\begin{deluxetable}{c c c c c c c c c}
\tablecaption{Cosmic variance for different surveys with constant
  comoving volume.  The first and second columns show the mean
  redshift and the redshift bin size, the third column gives the root
  comic variance for dark matter, and the last six columns give the
  root cosmic variance for galaxies in the indicated stellar mass
  intervals.}

\startdata
\hline
\hline\\
~ & ~ & ~ & \multicolumn{6}{c}{$\sigma_{gg}$ for $\log(m_*/\msun)\pm0.25$}\\
$z_{\rm min}$ & $z_{\rm max}$ & $\sigma_{dm}$ & ~ & ~ & ~ & ~ & ~\\
~&~&~&8.75&9.25&9.75&10.25&10.75&11.25\\

\cutinhead{GOODS $10'\times16'$}\\
0.00 & 1.12 & 0.081 & 0.099 & 0.103 & 0.107 & 0.116 & 0.126 & 0.168\\
1.12 & 1.58 & 0.084 & 0.133 & 0.143 & 0.148 & 0.163 & 0.194 & 0.291\\
1.58 & 1.99 & 0.078 & 0.149 & 0.163 & 0.174 & 0.192 & 0.241 & 0.374\\
1.99 & 2.39 & 0.073 & 0.165 & 0.183 & 0.204 & 0.225 & 0.295 & 0.464\\
2.39 & 2.78 & 0.069 & 0.186 & 0.208 & 0.243 & 0.267 & 0.365 & 0.573\\
2.78 & 3.17 & 0.065 & 0.209 & 0.235 & 0.290 & 0.316 & 0.446 & 0.696\\
3.17 & 3.58 & 0.060 & 0.231 & 0.262 & 0.339 & 0.368 & 0.534 & 0.823\\
3.58 & 4.00 & 0.056 & 0.260 & 0.296 & 0.404 & 0.436 & 0.647 & 0.983\\

\cutinhead{GEMS $28'\times28'$}\\
0.00 & 1.12 & 0.063 & 0.077 & 0.080 & 0.083 & 0.091 & 0.098 & 0.130\\
1.12 & 1.58 & 0.060 & 0.096 & 0.103 & 0.107 & 0.118 & 0.140 & 0.210\\
1.58 & 1.99 & 0.055 & 0.105 & 0.114 & 0.122 & 0.135 & 0.169 & 0.262\\
1.99 & 2.39 & 0.050 & 0.114 & 0.126 & 0.140 & 0.155 & 0.203 & 0.319\\
2.39 & 2.78 & 0.047 & 0.126 & 0.141 & 0.165 & 0.181 & 0.247 & 0.389\\
2.78 & 3.17 & 0.043 & 0.140 & 0.158 & 0.194 & 0.212 & 0.299 & 0.466\\
3.17 & 3.58 & 0.040 & 0.153 & 0.173 & 0.225 & 0.244 & 0.354 & 0.545\\
3.58 & 4.00 & 0.037 & 0.171 & 0.194 & 0.265 & 0.286 & 0.425 & 0.645\\

\cutinhead{COSMOS $84'\times84'$}\\
0.00 & 1.12 & 0.036 & 0.044 & 0.046 & 0.048 & 0.052 & 0.057 & 0.075\\
1.12 & 1.58 & 0.030 & 0.047 & 0.051 & 0.053 & 0.058 & 0.069 & 0.104\\
1.58 & 1.99 & 0.026 & 0.049 & 0.054 & 0.058 & 0.064 & 0.080 & 0.124\\
1.99 & 2.39 & 0.023 & 0.052 & 0.057 & 0.064 & 0.070 & 0.093 & 0.145\\
2.39 & 2.78 & 0.021 & 0.056 & 0.063 & 0.074 & 0.081 & 0.110 & 0.173\\
2.78 & 3.17 & 0.019 & 0.061 & 0.069 & 0.085 & 0.093 & 0.131 & 0.204\\
3.17 & 3.58 & 0.017 & 0.066 & 0.075 & 0.097 & 0.105 & 0.153 & 0.235\\
3.58 & 4.00 & 0.016 & 0.073 & 0.083 & 0.113 & 0.122 & 0.181 & 0.275\\

\enddata
\label{t:cvsurveyvol}
\end{deluxetable} 

\subsection{Cosmic variance for different redshifts}
\label{sec:redshift}

We would expect the cosmic variance of dark matter to decrease
monotonically with increasing redshift, because it depends on the
clustering which is lower for high redshift. However, we see, that for
the UDF $\sigma_{dm}$ increases, reaches a maximum and then decreases
with increasing redshift, while for the other surveys $\sigma_{dm}$
decreases with increasing redshift. The reason for this is that
although the different redshift bins have the same comoving volume,
their geometry is not the same. This is due to the pencil beam
geometry for which the volume of the low redshift bins are long but
narrow, while the higher redshift bins become shorter and wider. Thus
with increasing redshift the ratio of the area and the depth of the
redshift bins increases. In the lower redshift bins the galaxies are
thus on average more separated than in the higher redshift bins. This
causes the cosmic variance to increase with redshift. At high redshift
and for very wide surveys the ratio decreases again with redshift as
the line of sight distance in a redshift bin becomes smaller than the
transverse distance.

Both effects, the decrease of $\sigma_{dm}$ with redshift due to lower
clustering amplitudes and the growth of $\sigma_{dm}$ because of the
geometry of the redshift bins, affect $\sigma_{dm}(z)$. The redshift
at which $\sigma_{dm}$ is maximal depends on the area of the
survey. For small areas (e.g. UDF) it is higher ($z_{max}\sim1.3$)
while for large areas (e.g.  COSMOS) it is lower ($z_{max}<0.2$). The
cosmic variance for the galaxy samples we consider, however, always
increases with redshift. This is due to the large biases of these
samples, which are always larger than unity. Since the bias increases
strongly with redshift, the cosmic variance for galaxies thus also
increases with redshift.

Figure \ref{fig:cvsurvey} plots the cosmic variance as a function of
redshift with redshift bins of constant volume. The four different
panels show $\sigma_{v}$ for dark matter and three stellar mass
bins. The lines are for different surveys. We see that except for the
UDF, the cosmic variance for dark matter decreases with redshift while
for all surveys the cosmic variance for galaxies increases with
redshift.

\subsection{Cosmic variance for different geometries}
\label{sec:ratio}

\begin{figure}
\psfig{figure=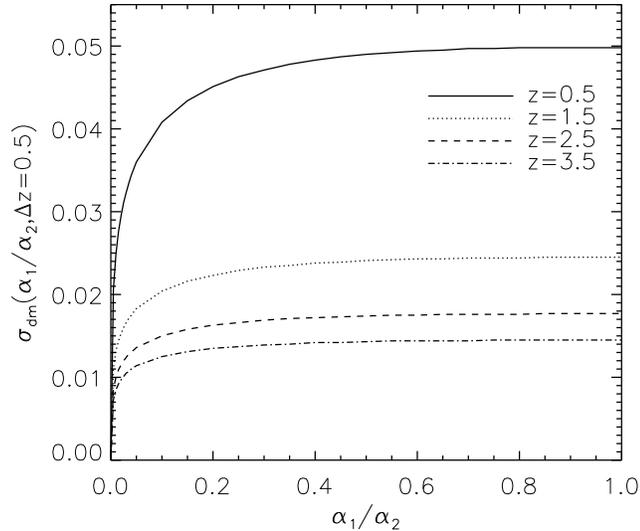,width=0.47\textwidth}
\caption{\scriptsize Effects of survey geometry on cosmic
  variance. The root cosmic variance of dark matter is plotted as a
  function of the ratio between the two observation angles
  $\alpha_1/\alpha_2$. The lines represent different redshift bins
  with a bin size of $\Delta z=1.0$.}
\label{fig:ratio}
\end{figure}

As we have shown, cosmic variance depends strongly on the geometry of
the survey. In section \ref{sec:redshift} we investigated the geometric
effect that arises because of the ratio between line-of-sight distance
and the transverse distance in the redshift bins. Now we investigate
the effects that arise due to different angular geometry on the
sky. For this we assume a survey with a fixed total area of $1.0{~\rm
  deg}^2$. We vary the ratio between the two observation angles
$\alpha_1/\alpha_2$ from $\alpha_1/\alpha_2\approx0$ to
$\alpha_1/\alpha_2=1$. This is done for several mean redshifts with a
bin size of $\Delta z=0.5$.

Figure \ref{fig:ratio} shows the geometry effects on the cosmic
variance. It plots $\sigma_{dm}$ as a function of the ratio between
the two observation angles. The result agrees with our explanation in
section \ref{sec:redshift}: for a low ratio the cosmic variance is
very low. If one increases the ratio the cosmic variance quickly
increases and reaches its maximum at $\alpha_1/\alpha_2=1$.  Here the
mean distance between the observed galaxies is the smallest, so that
many galaxies are likely to be correlated, resulting in large cosmic
variance.

This effect can also be seen by comparing the cosmic variance for GEMS
and EGS at a fixed redshift bin.  We find that although the area of
the GEMS field is larger than that of the EGS, GEMS has a higher
cosmic variance.  However, we note that unless the ratio between the
observation angles is smaller than $\sim20\%$, this effect is
small. In agreement with \citet{newman2002}, we find that the size of
the survey area is much more important than the axis ratio.

\section{Conclusions}
\label{sec:conclusions}

Inferences about galaxy number densities and related quantities such
as luminosity and stellar mass functions are subject to uncertainties
due to cosmic variance. It can be a significant source of uncertainty,
especially for surveys which cover only small areas and for massive
high-redshift galaxies. We have derived a simple recipe to compute
cosmic variance for five surveys as a function of mean redshift
$\bar{z}$, redshift bin size $\Delta z$ and the stellar mass of the
galaxy population $m_*$.

For this recipe we first calculated the dark matter cosmic variance as
a function of mean redshift and for a reference bin size of $\Delta
z_{\rm red}=0.2$. This was done by integrating the dark matter
correlation function obtained from CDM theory for a pencil beam
geometry in a redshift bin. We provide fitting functions for five
surveys (UDF, GOODS, GEMS, AEGIS, COSMOS). We then
showed that the dependence of cosmic variance on bin size $\Delta z$
is nearly independent of mean redshift and survey geometry. This means
that cosmic variance for a different bin size can be estimated from
the value computed with the reference bin size using a simple
conversion factor. We have used the galaxy bias predictions by
\citet{moster2009} (for six stellar mass bins and six thresholds) and
the dark matter cosmic variance to compute the cosmic variance for
galaxies. We also presented a formula to compute the overall
cosmic variance for a volume-weighted average of multiple separated
fields.

Applying this recipe to GOODS, GEMS and COSMOS at $\bar{z}=2$ and
$\Delta z=0.5$, the relative cosmic variance for galaxies with
$m_*>10^{11}\msun$ ($m_*\sim10^{10}\msun$) is $\sim 38\%$ ($19\%$) for
GOODS, while it is $\sim 27\%$ ($13\%$) for GEMS and $\sim 12\%$
($6\%$) for COSMOS.  At $\bar{z}=3.5$ and $\Delta z=0.5$ we found a
relative cosmic variance for galaxies with $m_*>10^{11}\msun$
($m_*\sim10^{10}\msun$) of $\sim 78\%$ ($34\%$) for GOODS, $\sim 51\%$
($22\%$) for GEMS and $\sim 21\%$ ($9\%$) for COSMOS.

We find that for all fields the cosmic variance is much more
significant for massive galaxies, which is due to the strong dependence
of the galaxy bias on stellar mass.  Our results imply that cosmic
variance is a significant source of uncertainty at $\bar{z}=2$ for
small fields and massive galaxies, while for larger fields and
intermediate mass galaxies, cosmic variance is less serious. At
$\bar{z}=3.5$ cosmic variance can become serious in small fields even
for intermediate mass galaxies. For massive galaxies even the widest
survey (COSMOS) has significant uncertainties due to cosmic variance.

\acknowledgments{
BPM thanks the Space Telescope Science Institute for hospitality during part of this work.
JAN acknowledges support from NSF grant AST-0806732.
}

\clearpage

\end{document}